\documentclass[preprint,amsmath,amssymb,aps,prl,showpacs, showkeys]{revtex4-1}

\usepackage{graphicx}
\usepackage{dcolumn}
\usepackage{bm}

\let\ltxRightarrow\Rightarrow
\usepackage{marvosym}
\let\Rightarrow\ltxRightarrow
\usepackage{wasysym}
\usepackage{amssymb}
\usepackage{color}
\definecolor{gray}{rgb}{0.5, 0.5, 0.5}
\usepackage{enumitem}

\newcommand{\I}{\ensuremath{\mathrm i}} 

\begin{document}

\title{Realizing Quantum Materials with Helium: \\
Helium films at ultralow temperatures, from strongly correlated atomically layered films to topological superfluidity
}


\author{J. Saunders$^*$}

\address{Department of Physics, Royal Holloway University of London,
Egham, Surrey TW20 0EX, UK\\
$^*$E-mail: j.saunders@rhul.ac.uk\\
www.royalholloway.ac.uk}

\begin{abstract}
This article provides an overview, primarily from an experimental perspective, of recent progress and future prospects in using helium to realize a range of quantum materials of generic interest, by ``top-down'' and ``bottom-up'' nanotechnology. We can grow model systems to realise new quantum states of matter, and explore key issues in condensed matter physics. In the language of cold atomic gases, two dimensional and confined $^3$He and $^4$He provide ``quantum simulators'', with the potential to uncover new emergent quantum states. These include: strictly 2D Fermi system with Mott-Hubbard transition; interacting coupled 2D fermion-boson system; heavy fermion quantum criticality; ideal 2D frustrated ferromagnetism; 2D quantum spin liquid; intertwined superfluid and density wave order with emergent large symmetry;
topological mesoscopic superfluidity (new materials and emergent excitations).


\begin{description}

\item[Keywords]{strongly correlated fermions; two dimensional; Kosterlitz-Thouless
superfluid transition: supersolid; intertwined order; quantum spin liquid;
topological superconductivity and superfluidity; chiral superfluid; Majorana
fermions.}
\vspace{0.5in}
\item[Published]{Now published in \textbf{Topological Phase Transitions
and New Developments}, Lars Brink, Mike Gunn, Jorge V Jos\'e, John Michael
Kosterlitz, Kok Khoo Phua, editors. World Scientific 2018, pp 165--196, 
https://doi.org/10.1142/9789813271340\_0012.
}
\end{description}

\end{abstract}

\maketitle

\section{1. Introduction\label{sec:introduction}}
Helium is unique: for both isotopes the combination of relatively weak van der Waals atomic attractions and strong zero point motion, due to small mass, leads to the stability of the bulk liquid phases down to absolute zero. These model systems of strongly correlated bosons ($^4$He) and fermions ($^3$He) have played a central role in the development of concepts in condensed matter physics. Superfluid $^4$He demonstrated the first BEC, albeit with condensate fraction strongly depleted by interactions, and the first macroscopic quantum state. Liquid $^3$He led to the development of Landau Fermi liquid theory---the standard model of strongly correlated fermions---with the striking \emph{prediction} of collisionless zero sound, a collective-mode of the Fermi surface that can be driven and detected ultrasonically. 

The topic of this workshop is topological phase transitions and topological quantum matter. Topological quantum matter has been classified \cite{Schnyder2008} and has developed into a concept of wide applicability~\cite{Qi2011,Hasan2010} with the discovery of topological insulators, and proposals of potential topological superconductors. Thus as an established topological superfluid $^3$He is clearly a system of significant contemporary importance~\cite{Mizushima2015,Mizushima2016}.

Superfluid $^3$He was the first discovered unconventional superconductor/superfluid (Nobel Prize 1996, 2003)\cite{Leggett1975b,Leggett2006,Vollhardt2013,Volovik2003}. It has $L=1$ (p-wave) and $S=1$ pairing, with a 9 (complex) component tensor order parameter allowing multiple superfluid phases. Normal liquid $^3$He is the paradigm Landau Fermi liquid, underpinning the ``standard model'' of strongly correlated quantum matter; there is no lattice and the Fermi surface is a perfect sphere. The high degree of symmetry, $\text{SO}(3)_\mathrm S \times \text{SO}(3)_\mathrm L \times \text{U(1)}\times \text{T}\times \text{P} $, of the normal state is a key simplifying factor in revealing the broken symmetries of the emergent topological superfluid phases. It is easier to study anisotropic superfluidity if the Fermi surface is isotropic. There are two main stable phases in bulk, which break the symmetry of the normal state in different ways. There is clear evidence of pairing by spin fluctuations~\cite{Anderson1973b}, extensively sought and discussed in heavy fermion superconductivity. Furthermore the importance of topology in momentum space, and the emergence of surface and edge excitations in these superfluids was discussed extensively in the relatively early literature~\cite{Volovik1988,Volovik1992,Volovik1992b,Volovik2009,Volovik2009b,Volovik2010}. 

The 2016 Nobel Prize reminds us of the power of $^4$He to experimentally test theoretical concepts, in this case through the observation of the topological phase transition in superfluid $^4$He films, as discussed elsewhere in this volume. (The fact that this was almost 40 years after the first detection of superfluid $^4$He film flow graphically illustrates the power of interplay between theory and experiment). The $^4$He film was grown on mylar, which is a heterogeneous substrate (the binding potential is non-uniform over the surface). The key features are that, beyond a $^4$He so-called ``dead-layer'' of localised atoms a uniform mobile film is created, the density of which can be tuned continuously. The vortex interactions are precisely of the required logarithmic form (avoiding screening effects that plagued the observation in a 2D superconductor).  And the experimental method of using a high $Q$ torsional oscillator to observe the decoupling of the film from the substrate at the superfluid transition yield both the real and imaginary parts of the response function to validate the finite-frequency theory. This provides one topical template to highlight the power of helium to test key ideas in condensed matter physics.

This article mainly discusses the work of our group, as reported at this symposium. It is not a comprehensive review. It is intended to provide a broadly accessible overview of our research on helium films.  

\section{2. Bottom up: strong correlation effects in two dimensional helium}

The approach is to use graphite as a substrate. Atoms are physisorbed onto this substrate in a controlled way. The resultant film is ``self-assembled'', driven by a combination of atomic interactions with graphite, helium interatomic interactions, and zero point energy.

Graphite is atomically flat so that helium films grown on its surface are atomically layered. This is established both experimentally and theoretically. The binding energy of a $^3$He atom to graphite is around 150\,K, while that of $^4$He is higher due to the larger mass and hence smaller zero point energy \cite{Godfrin1995}.

As the helium film is grown the complete first layer forms a compressed incommensurate solid on a triangular lattice. Atoms added to the second layer are subject to a periodic potential due to these first layer atoms. Typically films up to eight atomic layers have been studied. A key tool in the study of 2D $^3$He is to pre-plate the graphite surface to effectively create a new substrate. This exploits the fact that $^3$He is more weakly bound than any other species. Before growing a $^3$He film, the substrate can be pre-plated, for example with: a solid monolayer of $^4$He; a solid bilayer of $^4$He; a solid HD bilayer.  We find that each composite substrate allows us to use the subsequently grown $^3$He film to model different physics. Thicker $^4$He films consist of a number of solid layers next to the substrate with a superfluid $^4$He overlayer. When $^3$He is adsorbed onto this surface, the superfluid $^4$He ``substrate'' plays an active role due to fermion-boson coupling.

These atomically layered helium films have allowed the study of a wide variety of phenomena:
\begin{itemize}
\item   the study of 2D Fermi systems to test of the applicability of Landau Fermi liquid theory in 2D;
\item       the density-tuned Mott transition of a $^3$He monolayer;
\item       Kondo-breakdown quantum criticality in a $^3$He bi-layer;
\item       frustrated magnetism in 2D; Heisenberg ferromagnet in 2D;
\item       quantum spin liquid in a $^3$He solid monolayer;
\item       Kosterlitz-Thouless transition in $^4$He films;
\item       intertwined superfluid and density wave order in $^4$He films -- a 2D supersolid.
\end{itemize}

\noindent

In practice exfoliated graphite is used as the adsorbate, to provide sufficient total surface area for measuring the properties of the adsorbed helium film. Typical specific surface areas range from $1-20~$m$^{2}$/gm, depending on preparation. The potential influence of finite platelet size, interconnectivity and orientation of platelets, residual surface disorder are taken into account in the analysis. In particular, the detection of superfluidity of a 2D Fermi system, that is of a $^3$He fluid monolayer, requires an adequately small level of surface disorder, to which p-wave superfluidity would be particularly sensitive. 

\begin{figure}[h]
\begin{center}
\includegraphics[width=2in]{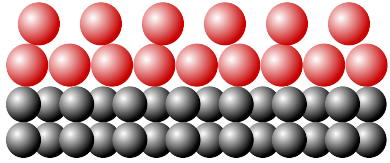}
\end{center}
\caption{Cartoon of an atomically layered helium film on the atomically flat surface of graphite. Helium atoms are represented by hard spheres. Shown is a two layer film. The complete first layer forms a compressed solid on a triangular lattice. In such films the total areal density (coverage) can be adjusted with precision.}
\label{fig:fig-1}
\end{figure}

\subsection{2.1 Mott-Hubbard transition}
 A monolayer of $^3$He on graphite preplated by a bilayer of HD shows, from heat capacity and magnetization measurements, a clear density-tuned effective mass-diverging Mott transition, at which the Landau Fermi liquid parameter $F_0^\mathrm a$ saturates \cite{Casey2003d}. This confirms the almost-localised fermion picture of strongly correlated fluid $^3$He, in which the strongly repulsive hard core repulsion is dominant~\cite{Vollhardt1984}. This system also allowed the measurement of non-analytic finite $T$ corrections to the heat capacity, which support the theoretical claim that, although anomalous, Fermi liquids can survive in 2D \cite{Chubukov2005}, in contrast to earlier ideas~\cite{Anderson1990}.

 A related case is the second layer of $^3$He on graphite (including the case where the first layer is replaced by $^4$He)~\cite{Lusher1991b}. Our recent NMR study~\cite{Arnold20xx} shows a relatively wide density range over which there is a quantum coexistence of fluid and solid, with no evidence for a hole-doped Mott insulator and associated Fermi surface reconstruction. In this case we argue that the $^3$He experiences a density tuned Wigner-Mott-Hubbard transition.  However, unresolved issues remain with behaviour at the very lowest temperatures $T\ll T_\mathrm F^*$, in terms of unexplained temperature dependences of the heat capacity and magnetization. Whether this behaviour is intrinsic, arising from fermionic correlations (for example formation of a fermionic flat-band)~\cite{ClarkKhodel2005}, or extrinsic, arising from residual surface disorder, may be tested by experiments using substrates of improved quality.

In both cases discussed above the Mott insulator is a strong candidate for the long-sought quantum spin liquid material, as discussed later. 

\begin{figure}[h]
\begin{center}
\includegraphics[width=2in]{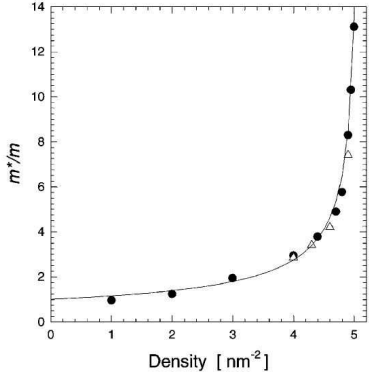}
\end{center}
\caption{Effective mass divergence at a density driven Mott transition, in a 2D $^3$He monolayer. The substrate is graphite, plated by a bilayer of HD \cite{Casey2003d}. The tuning parameter is the $^3$He surface density. The effective mass is determined from the heat capacity. The effective mass inferred from magnetization data are consistent, within the framework of the almost localized fermion model, for which $(1+F_0^\mathrm a)$ saturates at a finite value of approximately $1/4$. 2D $^3$He is almost localised \emph{not} almost ferromagnetic. The Mott insulator is a strong candidate to be the elusive quantum spin liquid.}
\label{fig:fig-2}
\end{figure}

\subsection{2.2 Heavy fermion physics and quantum criticality in a $^3$He bilayer}
When, on the other hand, $^3$He is grown on graphite plated by a bilayer of solid $^4$He, yet new strongly correlated physics emerges. The fluid $^3$He bi-layer behaves as a heavy fermion system~\cite{Neumann2007}. The lower layer (L1) plays the role of the f-fermions and the second layer (L2) is analogous to the mobile conduction electrons. The two layers are hybridized by a Kondo interaction: in this case exchange of atoms between the two layers. This is tuned by the density of the upper layer. A density-tuned quantum critical point (QCP) is found at which the effective mass diverges. This appears to fall into the class of orbital-selective Mott transition~\cite{Vojta2010}, with a Kondo-breakdown QCP~\cite{Benlagra2008,Pepin2008,Rancon-Schweiger2011,Beach2011,Sen2016}, at which the effective mass diverges. Beyond this QCP, layer L1 is localised and layer L2 is itinerant, consisting of weakly-interacting 2D fermions. The frustrated magnetism of atomic ring exchange plays a role in L1.  Approach to the QCP is intercepted by a magnetic instability, which it is believed is triggered when the ferromagnetic intralayer exchange in L1 dominates the interlayer Kondo coupling~\cite{Neumann20xx,Werner2014}.

\subsection{2.3 Superfluidity of atomically layered $^4$He films}
If a $^4$He film is grown on graphite, superfluidity is observed in the third and subsequent layers~\cite{Crowell1996}. The first two layers are solid $^4$He. Anomalous properties in the second layer, interpreted as a 2D "supersolid"  are discussed in Section~2.6. A related system is $^4$He on graphite plated by HD~\cite{Nyeki1998}. Superfluidity in these atomically layered films shows a Kosterlitz-Thouless (KT) transition. The KT transition, when measured using a torsional oscillator at finite frequency, has a somewhat rounded ``jump'' in the superfluid density accompanied by a peak in dissipation. These are inferred from the period shift and quality factor, which determine the real and imaginary parts of the vortex response function, respectively.  A parametric Cole-Cole plot of the real \emph{vs}. imaginary parts of the superfluid response function provides a characteristic fingerprint of the transition. Indeed the data for a range of $^4$He coverages, and for a variety of surface preplatings collapse well onto a ``universal'' contour~\cite{Bowley1998}, albeit with small quantitative discrepancies relative to the theoretically predicted contour.

\begin{figure}[h]
\begin{center}
\includegraphics[width=3in]{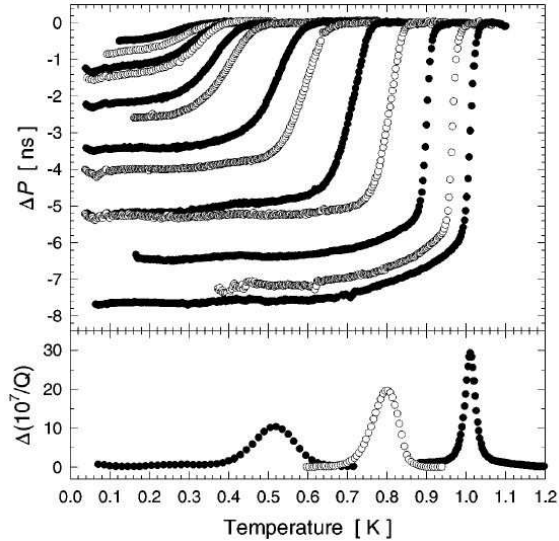}
\end{center}
\caption{Kosterlitz-Thouless superfluid transitions of $^4$He films on graphite plated by a trilayer of HD~\cite{Nyeki1998}. $^4$He coverages range from $7.0$ to $13.1$~nm$^{-2}$. Similar results are found with HD bilayer preplating. A parametric plot of the real vs. imaginary parts of the superfluid response, for both preplatings, collapse onto a ``universal'' curve~\cite{Bowley1998}.   }
\label{fig:fig-3}
\end{figure}

There is a further important feature of the growth of these atomically layered $^4$He films. As each layer forms, it undergoes spinodal decomposition,  at layer densities below around 4\,nm$^{-2}$. In other words each 2D fluid layer will self-condense (liquid-gas instability in 2D) \cite{Gordill01998,Pierce1999} forming puddles with a self-bound density 4\,nm$^{-2}$. This effect is inhibited in $^4$He films on heterogeneous substrates. We draw the reader's attention to an unexplained observation that we find particularly striking. This is the suppression of superfluidity approaching the spinodal point from higher densities, observed both with and without HD pre-plating, that suggests an intrinsic  mechanism for the suppression of 2D superfluidity, as yet theoretically unexplained. We conjecture that this may either arise from the strong periodic potential of the first solid $^4$He layer atoms, or from the influence of the proximity of the spinodal point on the vortices and their dynamics.

\subsection{2.4 $^3$He 2D Fermi liquid}
Nevertheless, the three layer film (superfluid $^4$He monolayer sitting on a solid $^4$He bilayer), and the four layer film (superfluid $^4$He bilayer sitting on a solid $^4$He) are relatively well understood. They appear to provide ideal ``substrates'' for the study 2D $^3$He. The study of these, so-called, helium mixture films has been the the subject of extensive prior work using heterogeneous substrates~\cite{Hallock1995,Hallock2000}. The use of atomically layered $^4$He films on graphite, as in the work descibed here, has significant advantages. These films are adsorbed on graphite, and so can be readily cooled into the microkelvin regime. Furthermore, the uniformity of the $^4$He film is of crucial importance, since it guarantees a uniformity of the fermion-boson coupling, which is enhanced by the graphite substrate~\cite{Yokoshi2003}. 

As $^3$He is added to the $^4$He film, a 2D $^3$He Fermi system of tuneable density is built on the single particle surface ground state for motion normal to the surface. We find that on the 3-layer $^4$He film, the $^3$He system shows a number of instabilities for coverages below 1\,nm$^{-2}$, with possible evidence for $^3$He dimer formation in a very low density component~\cite{Waterworth20yy}. At higher $^3$He densities above 1\,nm$^{-2}$ the superfluidity of the $^4$He film is suppressed and quenched for coverages above around 4\,nm$^{-2}$. On the other hand, for the 4-layer film the instability region at low $^3$He coverages is significantly reduced, and a single Fermi system can be studied up to 4.5\,nm$^{-2}$, at which (in the most na\"ive picture) it becomes energetically favorable for two (ground and first excited) surface normal states  to become occupied~\cite{Dann1999}, see Fig.~4. In practice these two sub-bands will be hybridized versions of those expected in a non-interacting ideal Fermi gas picture.

Over the coverage range at which the $^3$He forms a single 2D Fermi system, the conditions are satisfied for the interactions in this 2D system to be strictly two dimensional. This is in contrast to 2D cold atom systems studied thus far, where interactions are 3D (eg. the system can be tuned through a Feshbach resonance, absent for 2D interactions). Thus $^3$He provides an interesting model system to explore interactions in 2D, of significant interest theoretically~\cite{Engelbrecht1992,Engelbrecht1992b,Chubukov1993,Chubukov1994,Miyake1983b,Kagan1994}.  This is important in the wider context of: 
two dimensional/interface superconductivity in metals~\cite{Qin2009}, incuding
FeSe/STO~\cite{Ge2015}, Li/graphene~\cite{Profeta2012,Ludbrook2015}; cold atoms, experiments on multiple quasi-2D layers~\cite{Sommer2012,Feld2011}. 

The $^3$He system allows a number of key questions to be addressed:

\begin{itemize}
\item Do Fermi liquids survive in 2D \cite{Anderson1990}?
\item What is the nature of the interactions, in particular are there anomalies associated with interactions in 2D? [Here the ability to tune density over a wide range is important].
\item Is there evidence of $^3$He--$^3$He interactions mediated by the $^4$He bosonic film? [These interactions, mediated by phonon/3rd sound-like excitations are expected to be (graphite) substrate enhanced. Fermion and boson modes should be hybridized~\cite{Yokoshi2003}].
\item Can we enter a low density regime in which Fermi gas theory applies~\cite{Abrikosov1958,Abrikosov1959}, and Landau parameters can be calculated by a single, density dependent, scattering parameter?

With regard to the last point, this is challenging because of the logarithimic dependence of the interaction parameter in 2D;
\begin{equation*}
g=-1/\ln(2E_\mathrm F/E_\mathrm b) \quad \text{or} \quad g=-1/2\ln(\sqrt{2}k_\mathrm F  a).
\end{equation*}
     
\item Under what conditions might the 2D superfluidity of a system of fermions in 2D be realised~\cite{Miyake1983b,Kurihara1983}?
\end{itemize}

\begin{figure}[h]
\begin{center}
\includegraphics[width=2.5in]{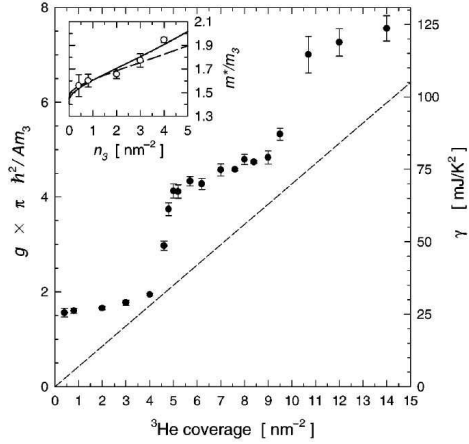}
\end{center}
\caption{Energy density of states, normalised by Fermi gas value, for $^3$He on a four layer $^4$He film on graphite. Below a $^3$He coverage of 4 nm$^{-2}$ the $^3$He forms a 2D Fermi system built on the ground state for motion along the surface normal. Steps indicate the population of excited surface-normal states (``multiple sub-band occupancy'').}
\label{fig:fig-4}
\end{figure}

So far the experiments have been to measure heat capacity  and nuclear magnetic susceptibility~\cite{Dann1999,Waterworth20yy} . The enhancement of the Pauli susceptibility relative to the ideal Fermi gas value has been determined by the highly sensitive low frequency SQUID NMR method~\cite{Arnold2014}, so far to temperatures as low as 200\,$\mu$K. NMR selectively measures the susceptibility of the Fermi system; there are no background corrections. In Landau Fermi liquid theory it is given by
\begin{equation*}
\frac{\chi}{\chi_0}=\frac{1+\frac{1}{2}F_1^\mathrm s}{1+F_0^\mathrm a}
\frac{m_\mathrm H}{m_3},
\end{equation*}
This is temperature independent for $T\ll T_\mathrm F^{**}$, where  
$T_\mathrm F^{**}=\frac{0.505(1+F_0^\mathrm a)}{m^*/m}n_3$ K\,nm$^{2}$. The Fermi gas susceptibility,  $\chi_0$,  is independent of $^3$He density in 2D. Thus the nuclear magnetic susceptibility provides a good measure of correlation effects to arbitrary low temperatures and for arbitrarily low surface density.
The enhancement of the coefficient of the linear in $T$ heat capacity is $\gamma /{\gamma _0} = (1 + {\textstyle{1 \over 2}}F_1^s)({m_\mathrm H}/{m_3}) $. In this way $F_0^\mathrm a$ can be determined with precision. The relative density dependence of $F_0^\mathrm a$ and $F_1^\mathrm s$ found experimentally \cite{Waterworth20xx}, can be compared with the predictions of microscopic theory \cite{Chubukov1994}, and we find that both s-wave and p-wave scattering are essential to explain results, such that back-scattering dominates. 

At somewhat higher density, with ``multiple sub-band occupancy'', we have two coupled 2D Fermi systems, of potential relevance to coupled 2D layers in cuprate superconductors.
\subsection{2.5 The second layer of helium films on graphite: 2D supersolid and quantum spin liquid}
In this section we focus on to two quantum states: a 2D supersolid (realised in a $^4$He atomic layer) and a quantum spin liquid (for which there is mounting evidence in a $^3$He atomic layer), and we speculate on the potential relationship between them.  As we shall see these two states are manifested at essentially the same helium coverage, occurring in the second atomic layer of helium on graphite, prior to the completion of that layer. All we do is replace bosons by fermions, with the same interactions, albeit different mass. We ask, might this fermion-boson ``correspondence'' be significant?

The realization of supersolid in cold atomic gases is a matter of active research \cite{Leonard2017,Li2017}. The realization of the quantum spin liquid in layered magnetic materials has been a long-term quest, with many ``candidates''~\cite{Balents2010,Balents2017}. It would be quite appealing if both of these highly entangled quantum ground states were found in two dimensional helium.

\subsection{2.6 Intertwined superfluid and density wave order: 2D supersolid\label{sect-supersolid}}
The 2D supersolid phase that we identify \cite{Nyeki2017,Nyeki2017b} is quite different from the putative supersolid phases proposed in the literature in the context of bulk solid $^4$He.  We argue that the state of intertwined superfluid and density wave order has a larger emergent symmetry~\cite{Fradkin2015}. It is a new quantum state of matter.

On the other hand the ``classic'' supersolid has a well-defined solid crystal structure with mobile vacancies constituting the superfluid component~\cite{Leggett1970,Balibar2010,Boninsegni2012,Hallock2015}.  In the case of bulk solid $^4$He it was proposed that these arise spontaneously \cite{Andreev1982}. Experiments using torsional oscillators to detect the mass decoupling associated with superfluidity of the solid, typically performed at kHz frequencies, must contend with a large visco-elastic response of the crystal \cite{Saunders2009}. There is no clear experimental evidence for superfluidity by this method. There is evidence for superfluidity at the cores of dislocations \cite{Vekhov2012,Shin2017}, as predicted theoretically~\cite{Boninsegni2007}.
All these proposals concern superfluidity associated with defects, coexisting with crystalline order.  

On the other hand, in the $^4$He layer, we propose a 2D supersolid as a new quantum material~\cite{Nyeki2017,Nyeki2017b}. This exotic quantum state would have both the rigidity of a solid, but also paradoxically be able to flow without resistance. To account for the observations, we propose a new quantum state in which density wave order and superfluid order are fully quantum entangled, represented by 
\begin{equation*}
\left|\Psi\right> = \exp\left(\alpha_0 b^\dagger_{\mathbf q=0}+\sum_{\mathbf G} \alpha_{\mathbf G} b^\dagger_\mathbf G \right)\left|0\right>,
\end{equation*}  
which describes a quasi-condensate at both zero momentum and finite momenta $\mathbf G_i$, the set of reciprocal lattice vectors of triangular lattice. A quasi-condensate at $\mathbf q=0$, in the presence of density wave order, necessarily implies a quasi-condensate at the set of wavevectors $\mathbf G_i$.

We hypothesize that the state vector freely explores the Bloch hyper-sphere, and hence varying relative occupation of condensates, as well as varying degrees of ``solid'' and ``superfluid'' order.  In other words, the Hamiltonian commutes with rotations transferring particles between the quasi-condensates. This is not a fragmented condensate \cite{Mueller2006}. It is a quantum material described by a macroscopic wave function (actually with power law correlations in space because we are in 2D), which is a Schr\"odinger-cat like state: the entire system can be both solid \emph{and} superfluid. Thus quantum mechanics resolves the central paradox of the supersolid, arising from our everyday intuition which tells us that the properties of a solid and a superfluid are contradictory. This is not a state of matter in which solid and superfluid simply coexist. It is an entangled quantum state in which density wave order and superfluid order are intertwined.

We briefly summarise the observations that lead to these proposals; a fully referenced discussion is provided elsewhere~\cite{Nyeki2017}. We measure the superfluid response of a bilayer $^4$He film adsorbed on exfoliated graphite with a torsional oscillator.
The first layer forms a compressed solid and, apart from some viscoelastic response, which can be corrected for, can be considered passive. Superfluid response occurs over a narrow range of film densities in the second layer close to its completion. The superfluid response shows no evidence of a KT transition (jump in superfluid density) and even shows no sharp onset. The superfluid density shows an anomalous temperature dependence; the leading order dependence in the $T \to 0$ limit is linear in $T$. We introduce the coverage/density dependent characteristic energy scale $\Delta(n)$, which governs both the temperature dependence of the normal fraction, and the $T=0$ superfluid fraction: 
\begin{equation*}
\frac{\rho_\mathrm s (T,n)}{\rho} = \frac{\Delta(n)}{T_0} f(T/\Delta(n)).
\end{equation*}
We can scale data to demonstrate quantum criticality. The quantum critical point, at which $\Delta(n)$  extrapolates to zero, is at layer completion.  The normal density in the $T \to 0$ limit has an  unusual linear in $T$ dependence. This is  accounted for by an \textit{ansatz} for the elementary excitation spectrum, in the spirit of Landau. The Landau prescription for calculating  $\rho_\mathrm n$ involves a momentum-weighted integral. The proposed spectrum has an extremely soft roton-like minimum, such that the roton gap is smaller than the minimum temperature explored so far experimentally (around 2\,mK). Nozi\`eres refers to the roton as the ``ghost of the Bragg spot''. The small roton gap also implies a strong peak in the structure factor via the Feynman-Cohen relation $E(q)=\hbar^2 q^2/2mS(q)$. Thus the inferred excitation spectrum implies density wave order. The \emph{ansatz} for the quasi-condensate wavefunction has both superfluid and density wave order. The fact that no Kosterlitz-Thouless transition is observed can be explained if the state is not such that the $\text{U}(1)$ symmetries of both orders are separately broken. Rather the two orders are fully intertwined, the ground state has $\text{SU}(N)$ symmetry, and vortices are not stable. This is a non-Abelian superfluid. It is important to note that these conclusions are drawn in the absence of knowledge of the system Hamiltonian. The starting point is not a model Hamiltonian that we simulate, but rather the emergent properties we experimentally observe.

The fact that this supersolid is observed in the second layer of helium on graphite is a manifestation of the highly quantum nature of the system. In bulk quantum solid helium~\cite{Andreev1982}, the atoms have high zero-point motion, the amplitude of which is a significant fraction of the lattice parameter. The overlap of neighbouring atomic wave-functions gives rise to atomic exchange, at a rate orders of magnitude smaller than the Debye frequency, and strongly dependent on lattice parameter (a function of pressure)~\cite{Guyer1971}, decreasing with molar volume as $V_\mathrm m^\gamma$  with $\gamma \sim 20$. Thus the atoms in solid helium move from site to site, and in solid $^3$He this manifests as an exchange interaction between the nuclear spins. Now, in the region of the second layer of helium on graphite that is presently of concern to us, the rate of inter-site atomic motion is significantly higher than in bulk solid because of the much lower density. This motion also has been detected through thermodynamic studies of $^3$He impurities introduced into the $^4$He second layer \cite{Ziouzia2003b,Nyeki20xx}.

\subsection{2.7 Quantum spin liquid}
We now turn to the potential realization of a quantum spin liquid phase in 2D solid $^3$He. Here the magnetism arises from the $^3$He nuclear spin, and is interrogated by NMR and heat capacity measurements~\cite{Godfrin1988b,Greywall1990b,Siqueira1997,Fukuyama2008,Collin2001}. In contrast to $^4$He, different phases (solid, fluid) can easily be distinguished through signatures in these thermodynamic probes.

The key feature of 2D solid $^3$He is that it forms a triangular lattice, and is therefore geometrically frustrated. Furthermore, the spin interactions are highly frustrated by competing atomic ring exchange~\cite{Roger1990}.  Ring exchange of an odd number of particles is ferromagnetic (FM), even is antiferromagnetic (AFM). The ring exchange interactions are strong in 2D, and significantly higher than in 3D solid helium, because of both high in-plane zero point motion, low density and zero point motion out of plane. Thouless \cite{Thouless1965} first proposed the effective spin Hamiltonian, in terms of permutation operators:
\begin{equation*}
\mathcal H = \sum_n (-1)^n J_n P_n \qquad \begin{array}{l}
P_2 = \frac{1}{2}(1+\boldsymbol\sigma_1\cdot\boldsymbol\sigma_2)  \\
P_3 = \frac{1}{2}(1+\boldsymbol\sigma_1\cdot\boldsymbol\sigma_2+\boldsymbol\sigma_2\cdot\boldsymbol\sigma_3+\boldsymbol\sigma_3\cdot\boldsymbol\sigma_1) \\
P_4 \text{ includes terms like } (\boldsymbol\sigma_1\cdot\boldsymbol\sigma_2)(\boldsymbol\sigma_3\cdot\boldsymbol\sigma_4) \\
\end{array}
\end{equation*}
The effective Heisenberg Hamiltonian  $J=J_2-2J_3$  is FM because three particle exchange dominates two particle exchange. This is a consequence of the fact that helium atoms are ``hard spheres''. 
\begin{figure}[h]
\begin{center}
\includegraphics[width=3in]{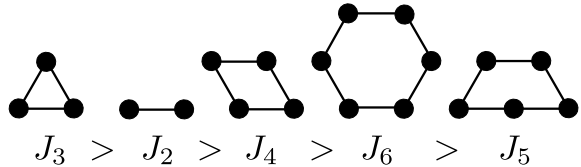}
\end{center}
\caption{Hierarchy of cyclic ring-exchange interactions in 2D $^3$He on a triangular lattice \cite{Roger1990,Roger1998}.}
\label{fig:fig-5}
\end{figure}

For simplicity, and the purposes of illustration, we truncate at 4 particle exchange. We refer to this two parameter model as the $J-J_4$ model. 
In principle these exchange parameters can be inferred from experiment, since the effective exchange parameters which enter the magnetic susceptibility, heat capacity, and spin wave velocity to leading order are different and take the form:
\begin{equation*}
\begin{split}
\text{Curie-Weiss constant} \qquad J_\chi &= -(J+3J_4) \qquad M=\frac{c}{T-\theta}
\quad \theta = 3J_\chi \\
\text{Spin wave velocity} \qquad J_\mathrm S &= -(J+4J_4)\\
\text{Heat capacity} \qquad J_c^2 &= (J+5J_4/2)^2 + 2J_4^2 \qquad C=\frac{9}{4}N k_\mathrm B \left(\frac{J_c^2}{T^2} \right).
\end{split}
\end{equation*}
These exchange parameters are tuned by the total film coverage, as shown in measurements of magnetization and heat capacity on the same samples \cite{Siqueira1997,Siqueira1996b}. 
The second layer with a fluid overlayer behaves as a 2D Heisenberg magnet with weak frustration: its properties are fully consistent with the Mermin-Wagner theorem \cite{Mermin1966} (there is no finite$-T$ phase transition). The temperature dependent spontaneous magnetism (observed by NMR in a weak magnetic field) at finite $T$ is accounted for by 2D spin-wave theory, with a small Zeeman gap. Perhaps this is the cleanest example of 2D ferromagnetism, with no complications from inter-layer coupling as in quasi-2D magnetic materials, and well characterised exchange.

 Our recent results suggest that the solid second layer is driven to ferromagnetism by the indirect RKKY interactions mediated by the fluid overlayer \cite{Waterworth20zz}. This is inferred from the evolution of the NMR lineshape (obtained using low frequency SQUID NMR), as it has been established that the third (fluid) overlayer initially forms in low density self-condensed puddles \cite{SatoFukuyama2012}. 
\begin{figure}[h]
\begin{center}
\includegraphics[width=4in]{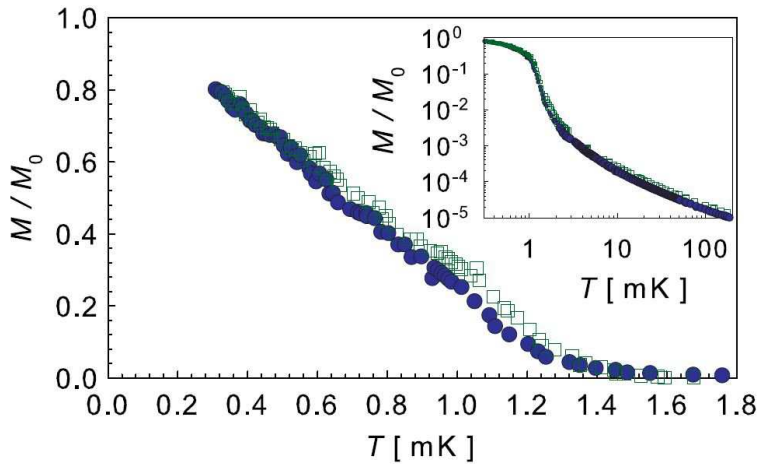}
\end{center}
\caption{Magnetization of the second $^3$He layer on graphite, in presence of a fluid overlayer, relative to its fully polarized saturation value at $T=0$. Results are consistent with Mermin-Wagner theorem for a 2D ferromagnet. Measurements in two magnetic fields show system is well described by 2D spin-wave theory. Results show that spin-wave spectrum is governed by a different exchange constant from that determining high-$T$ magnetization (see text), as a result of weak frustration~\cite{Casey2013}. }
\label{fig:fig-6}
\end{figure}

We now turn to the three physical $^3$He-on-graphite systems in which the quantum spin liquid (QSL) may manifest:
\begin{enumerate}[label=(\roman*)]
 
\item The second layer of $^3$He on graphite, where the first layer is $^3$He. In this case the first layer of $^3$He is a compressed solid on a triangular lattice [confirmed by neutron scattering]\cite{Roger1998}, paramagnetic, with very weak exchange interaction with the second layer. The coupled magnetism of the first and second layer is a complication. However the fact that the first layer is a weakly interacting ``spectator'' of the putative QSL in the second layer, may prove to be advantageous.
\item  A monolayer $^3$He on graphite, preplated by a solid monolayer of $^4$He. This system is very closely related to (i). However the paramagnetic $^3$He first layer is replaced with non-magnetic $^4$He. The density of the close packed $^4$He first layer triangular lattice is about 5\% higher than the $^3$He first layer. Given this close correspondence we will refer to this system also as ``the second layer of $^3$He on graphite''.
\item  A monolayer of $^3$He on graphite, preplated by a solid bilayer of HD~\cite{Siqueira1993,Casey1998b}.
\end{enumerate}
In all of the above cases, the putative QSL is a 2D solid on a triangular lattice, at the border of a density tuned Mott-Hubbard transition. This is a spin $\frac{1}{2}$ system [$^3$He nuclear spin]. As well as the geometrical frustration of the triangular lattice, there is strong frustration due to competing atomic ring exchange interactions. The magnetization is directly, and selectively, measurable by NMR. The high temperature magnetism shows that the system has an antiferromagnetic character. Magnetization measurements on both system (ii) and system (iii) support a gapless spin liquid~\cite{Masutomi2004}. In our recent work on system (ii) we find that the low temperature magnetism is consistent with a Pauli susceptibility \cite{Arnold20xx}, as expected for a gapless spin liquid, with a characteristic energy scale of a few hundred {$\mu$}K. 

Our current belief is that system (ii) and (iii) reflect a different balance between the periodic potential of the solid underlayer (HD bilayer or $^4$He) on a triangular lattice and intralayer $^3$He interactions. The HD bilayer is of significantly lower density than the $^4$He first layer. It shows a Mott-Hubbard transition into a 4/7 or 7/12 triangular superlattice phase. The results for system (ii) are more consistent with a density wave instability in the $^3$He layer. In system (iii) we have shown that exchange in the Mott insulator is much stronger than in system (ii)~\cite{Siqueira1993}. This is understood in terms of the lower density. Therefore a monolayer of $^3$He on graphite, preplated by a solid bilayer of HD may be the most promising for demonstrating quantum spin liquid behaviour.  

We can also use the information on ring exchange processes from the frustrated magnetism of $^3$He system (ii) to estimate the characteristic energy scale governing atomic mobility in the $^4$He (supersolid) case. Now all ring exchange processes add to translate a $^4$He atom between sites, so we have $t=J_2+2J_3+4J_4+\dots$\,. Using the values calculated by PIMC we find $t \sim 200$~mK, comparable to the temperature at which a superfluid response is detected.

\section{3. Top down: topological mesoscopic superfluidity}
\subsection{3.1 Motivation}
In contrast with the ``self-assembled'' films discussed in section $2$,  here we use nanofabrication methods to define a cavity. In the simplest case we have a thin slab geometry, into which helium is admitted through a fill line. This can be thought of as a film, of thickness precisely defined by the height of the cavity, with equivalent upper and lower surfaces. This strategy is particularly suitable for the study of superfluid $^3$He, where the diameter of the Cooper pair $\xi_0$ at zero pressure is around 80\,nm. [Stabilization of van der Waals films of such thickness is tricky in the face of competing effects of surface tension and gravity. Moreover it is not easy to precisely determine the film thickness of such a ``bottom-up'' thick film]. So far cavities of height $D$ in the range 1000 to 100\,nm have been studied, $k_F^{ - 1} \ll D < 10{\xi _0}$. While this is far away from the strictly 2D limit $k_\mathrm F^{-1}\sim D\ll\xi_0$, it should allow that limit to be approached in a controlled way in future experiments. In our approach, the thickness of the ``film'' formed by helium in the cavity is well defined by the confining geometry. Moreover for fixed cavity height the effective confinement $\xi_0/D$ is tuneable by pressure, since $\xi_0=h\nu_\mathrm F/2\pi k_\mathrm B T_\mathrm c$.  

Our motivation is the study of topological superfluid $^3$He under confinement, as a model system for topological superconductivity.  The topological classification of condensed matter systems has now become established as a key general principle to understand, predict and design new states of matter. There are several \emph{candidate} topological superconductors, discussed elsewhere in this volume. Here we exploit superfluid $^3$He, where the topological classification is firm. Under confinement the goal is to use it as a model system for the investigation of emergent surface excitations.

The gift of Nature is that the p-wave superfluid $^3$He supports both a time reversal invariant (TRI) phase ($^3$He-B) and a chiral phase ($^3$He-A). The richness of the order parameter of superfluid $^3$He also allows new potential phases (\emph{new ``materials''}), which should be stabilised by confinement. The most striking feature in topological quantum matter is the surface and edge excitations which emerge through bulk-surface/edge correspondence. In a topological superfluid these are (TRI superfluid) Majorana or (chiral superfluid) Weyl fermions. It is the understanding, detection, characterization and possible quantum control of these new particles, which is the key challenge. In our neutral topological ``superconductor'', the topological invariant is a property of the emergent superfluid order. Thus the ``protection'' of the surface/edge states is expected to be influenced by a subtle interplay between broken symmetry and topology~\cite{Mizushima2015,Mizushima2016}. 

This approach  opens the way to the study of $^3$He is studied in hybrid nano-structures, ``topological mesoscopic superfluidity''. $^3$He allows a degree of control that is unprecedented. Confinement (for example in a cavity of height $D$) is the key new control parameter and is used to stabilise distinct superfluid phases or normal phase, and the interfaces between them, with exquisite control over interface quality, the elimination of disorder, and flexibility of geometry.  The $^3$He hybrid mesoscopic structure can built from a range of ``materials'', with $^3$He tuned into a particular state by the scale and sculpture of confinement. Each superfluid phase has different symmetry, topology and surface/edge excitations, with further control over excitations at engineered intra-fluid interfaces. The helium-cavity surface interfaces are also of high quality: they are atomically flat and, as has been shown, the surface scattering is tuneable \textit{in situ}, by a $^4$He surface film~\cite{Murakawa2009,Okuda2012,Heikkinen20xx}. The flexibility this approach offers provides a clear opportunity to reveal edge, surface and interface states in a model topological superfluid.

\subsection{3.2 Topological superfluid $^3$He under engineered nanoscale confinement:
Methods}

The first experiments \cite{Levitin2013} required a number of technical breakthroughs: to confine $^3$He in a nanofabricated rectangular cavity of thickness comparable to the coherence length, including a precise characterisation of the geometry and surfaces of cavity; to cool the sample to well below the superfluid transition temperature; to ``fingerprint'' the superfluid order parameter by developing an NMR spectrometer of unprecedented sensitivity \cite{Levitin2007}; to measure superfluid density of such a small sample with a torsion pendulum.

The first generation of cells (fabricated by anodically bonding Hoya SD-2 glass and Silicon)~\cite{Dimov2010}, were of typical area 1\,cm$^2$, with a cavity height of 700\,nm. We subsequently developed all-silicon structures, direct wafer bonded, with cavity heights (in the range 50 to 300\,nm) precisely defined by an array of posts, and modelled successfully by finite-element techniques. The $^3$He confined within this cavity is cooled through the $^3$He in the metallic fill line. This relies on the fact that the thermal conductivity of normal liquid $^3$He at 1\,mK is high, comparable to that of high quality copper. The fill line links the sample to a silver sinter heat exchanger mounted on a cold-plate, which is thermally linked to the nuclear stage.
\begin{figure}[h]
\begin{center}
\includegraphics[width=3in]{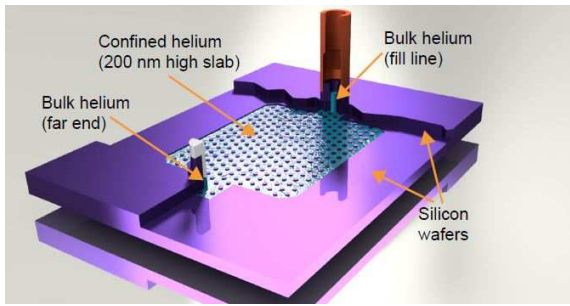}
\end{center}
\caption{Artist's impression of an all-silicon nanofluidic cell. Cavity height is $200$~nm, defined by array of silicon posts. Small volumes of bulk helium act as markers to compare with response of helium confined in cavity.}
\label{fig:fig-7}
\end{figure}

NMR plays a central role in the study of superfluid $^3$He.  The $^3$He nuclear spins are the spin degrees of freedom of the Cooper pairs; NMR directly interrogates the Cooper pairs, and through the coherent dipole interaction is used to ``fingerprint'' the superfluid order parameter (``pair-wavefunction''), in a direct and non-invasive way \cite{Leggett1974,Leggett1975b}. In the absence of Meissner effect and skin depth effects (present in superconductors) it can be used to probe the entire sample. This is very different from NMR studies of other quantum materials, where the nuclei are essentially spectators, hyperfine coupled to the strongly correlated electron system, and the focus is on Knight shift and $T_1$. The power of NMR was graphically illustrated by the rapidity with which the order parameters of superfluid $^3$He phases were established, following initial discovery \cite{Leggett1973,Leggett1974,Leggett1975b}. This should be contrasted with unconventional metallic and heavy fermion superconductors, where the complexity of the Fermi surface adds further to the problem.

While NMR thus provides, in principle, the method to directly determine the influence of confinement on the order parameter, including the stabilization of new phases, the experimental challenge is signal sensitivity of the NMR spectrometer. To address this question we developed a SQUID NMR method~\cite{Levitin2007}. The SQUID NMR technique allows measurements on a single well-characterized $^3$He slab. Previously confinement was achieved by immersing a stack of typically 1000 mylar sheets separated by dispersed polyamide spheres~\cite{Freeman1988,Freeman1990}. Unfortunately, this inevitably results in non-uniformity of confinement, which is undesirable since confinement is the key control parameter. The  presence of significant signals from surrounding bulk liquid also can complicate the discrimination of the NMR response of the confined sample.

The flexibility of nanofabrication allows the introduction of small volumes of ``bulk'' superfluid $^3$He, both near the fill line entrance to the cavity and at the far end of the cavity. These ``bulk-markers'', with well defined geometry, can be resolved from the cavity NMR signal, by simple NMR imaging techniques (``zeugmatography'' -- involving the applications of magnetic field gradients for spatial resolution). The ``bulk-markers''define the bulk-liquid superfluid transition temperature, and show that temperature gradients across the cavity are insignificant.

\subsection{3.3 Profound modification of the superfluid phase diagram due to confinement}

The relative stability of the A and B phases is strongly influenced by confinement. This arises because all three components of the orbital triplet are present in the B-phase, and the component with $l_z=0$ is most strongly suppressed by surface scattering. 
\begin{equation*}
\boldsymbol\Delta(\mathbf p)=\left[\Delta_\parallel(-\hat{p}_x+\I \hat{p}_y)\left|\uparrow\uparrow\right> +\Delta_\parallel(\hat{p}_x+\I \hat{p}_y)\left|\downarrow\downarrow\right>
+\Delta_\perp \hat{p}_z(\left|\uparrow\downarrow+\downarrow\uparrow\right>)
\right]\\.
\end{equation*}
On the other hand, in the A-phase, all pairs have the same orientation of their angular momentum,
\begin{equation*}
\boldsymbol\Delta(\mathbf p)= \Delta(\hat{p}_x+\I \hat{p}_y)\left(\left|\uparrow\uparrow\right>
+\left|\downarrow\downarrow\right>\right)
\end{equation*}
Under confinement the orbital angular momentum orients perpendicular to the wall. As a result the A-phase, which in bulk is only stable at high pressures in zero magnetic field, is favoured at low pressures, where the effective confinement parameter $D/\xi_0$ is smallest. In a 700\,nm cavity the A phase is stabilised at $T=0$ from zero up to around 2\,bar.~\cite{Levitin2013}
\begin{figure}[h]
\begin{center}
\includegraphics[width=3in]{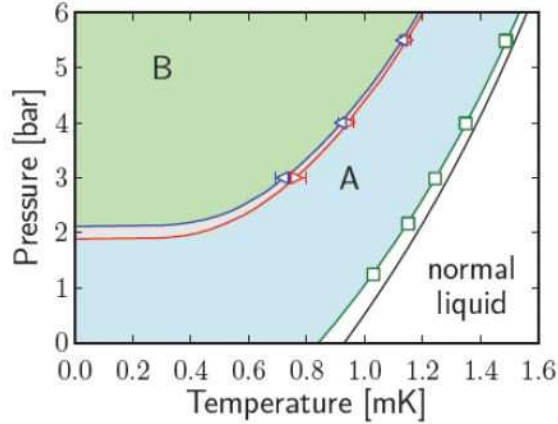}
\end{center}
\caption{The phase diagram of superfluid $^3$He confined in a slab-like cavity of height 700 nm \cite{Levitin2013}.
A-phase is stabilized at low pressure. Phase diagram shown is for diffusely scattering walls. In this case there is also a pressure dependent suppression of slab $T_c$ relative to bulk (solid black line).}
\label{fig:fig-8}
\end{figure}

Consistent results for the A-B transition line were found in NMR experiments on 700\,nm and 1100\,nm cavities \cite{Levitin2013,Levitin2019} and torsional oscillator experiments on a 1080\,nm cavity~\cite{Zhelev2017}. The critical value $D/\xi(T_\mathrm{AB})$ is pressure dependent and differs at all pressures from the prediction of weak coupling theory. The conclusion from these observations is that strong coupling corrections persist to zero pressure and are temperature dependent at each pressure. [Strong coupling refers to the dependence of the pairing interaction on the superfluid order parameter \cite{Sauls1981}]. 

As the cavity height is reduced,the expectation is that the A-phase will be stable over a progressively wider range of pressure. We expect that a re-entrant bubble (in the $p-T$ plane) of planar distorted B-phase will shrink and eventually disappear on reducing the cavity height: the details will depend on strong coupling effects and the influence of confinement. The superfluid phase under strong confinement $D\sim \xi_0$ is of great interest. These require control over the surface scattering conditions, which we now discuss.  
\subsubsection{3.3.1 Tuning the surface scattering}
Since the discovery of superconductivity in heavy fermion metals and oxide materials the majority of emerging superconducting materials exhibit unconventional, non-s-wave, pairing. In contrast to s-wave superconductors, they are extremely sensitive to non-magnetic defect, and surface scattering. Superfluid $^3$He is naturally defect-free (although there is a significant body of work investigating the influence of introducing disorder by immersing the superfluid in silica-aerogels).  
\begin{figure}[h]
\begin{center}
\includegraphics[width=4in]{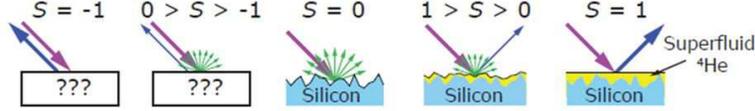}
\end{center}
\caption{Cartoon of different surface scattering conditions: from specular ($S=1$) through fully diffuse ($S=0$) to fully retro-reflecting $S=-1$. Eliminating the solid $^3$He boundary layer, by plating the surface with solid $^4$He, results in diffuse scattering. Adding further $^4$He to create a superfluid layer results in specular scattering. Creating the maximally pair-breaking retro-reflecting surface will require a specially engineered surface profile. }
\label{fig:fig-9}
\end{figure} 
In our experiments under confinement, the distinguishing feature is that we just have surface scattering, which is \emph{tuneable}; the surface scattering depends strongly on the helium surface boundary layer. Theoretically the calculation of surface suppression of components of the gap is made using quasiclassical theory; surface scattering is introduced phenomenologically~\cite{Vorontsov2003}. The limits are retroreflection, (random) diffuse, specular: specularity can be varied continuously between these limits. We have demonstrated the ability to tune surface quasiparticle scattering from magnetic, to non-magnetic diffuse, to specular, by precise measurements of $T_\mathrm c$ and gap suppression using NMR on a single cavity of height 200\,nm \cite{Heikkinen20xx}. The chiral A-phase is found to be stable at all pressures studied, $0-5.5$ bar.
Quasiclassical theory provides a consistent desciption of both the measured $T_\mathrm c$ suppression and gap suppression, in the non-magnetic scattering case.

With diffusively scattering surfaces p-wave superfluidity is suppressed for $D<\xi_0$. For perfectly specular surfaces both $T_\mathrm c$ and gap suppression are eliminated for order parameters with only $l_z=\pm 1$  components. These are the chiral A-phase and the TRI planar phase, which are degenerate in the weak coupling limit. Experiments are in progress, at the time of writing, to establish the phase diagram in $100$~nm cavities. In part they rely on the ability, demonstrated in the experiments on a $200$~nm cavity, to reproduce close to specular surface scattering conditions.

\subsection{3.4 Planar distorted B-phase}
In bulk the energy gap of the B-phase is isotropic. Under confinement the planar distorted a phase develops a strong gap anisotropy along $\mathbf o=\pm \mathbf z$, where $\mathbf z$ is the surface normal of the cavity . This results in a strong susceptibility anisotropy along $\mathbf w$, a vector rotated with respect to $\mathbf o$ by a rotation matrix $\mathbf R$, which rotates the spin relative to the orbital coordinates of the pair, changing only the dipole interaction. The Zeeman interaction is minimised for $\mathbf w=\pm \mathbf H$. This gives rise, in sufficiently strong magnetic fields, along $\mathbf z$, to two orientations with different dipole energies and hence different NMR frequency shift. For $\mathbf o=\mathbf z$ the dipole energy is minimised and the frequency shift is positive, while for $\mathbf o=-\mathbf z$ the frequency shift is negative. This latter metastable state is formed stochastically into the planar distorted B-phase from A-phase. It can be eliminated by growing the B phase in zero magnetic field.  Domain walls between these two orientations support surface states, including fermionic zero modes. We have proposed that they can be pinned at controlled sites on the surface and fused by manipulating the magnetic field \cite{Levitin2013b}. This is potentially a way of manipulating Majorana zero modes. These domain walls are ``soft'', of characteristic width $\xi_H\sim~10~\mu$m, the dipole length, and significantly wider than those referred to in the next section:
\begin{equation*}
\xi_H=\xi_0\sqrt{N_\mathrm F \Delta_\parallel^2 / \Delta \chi H^2}.
\end{equation*}
Under confinement the gap is spatially-dependent, due to wall boundary conditions. By measuring the NMR frequency shift as a function of tipping angle it is possible to measure the following spatial averages of the gap across the cavity:
\begin{equation*}
 \quad \bar{q} 
=\frac{\langle\Delta_\parallel(\mathbf r) \Delta_\perp(\mathbf r)\rangle}{\langle\Delta_\parallel^2(\mathbf r)\rangle}, \quad 
\bar{Q}^2 = \frac{\langle\Delta_\perp^2(\mathbf r)\rangle}{\langle\Delta_\parallel^2(\mathbf r)\rangle}.
\end{equation*} 
These can be compared with with predictions of the gap distortion from microscopic theory~\cite{Levitin2013b,Levitin2014}.
\begin{figure}[h]
\begin{center}
\includegraphics[width=4in]{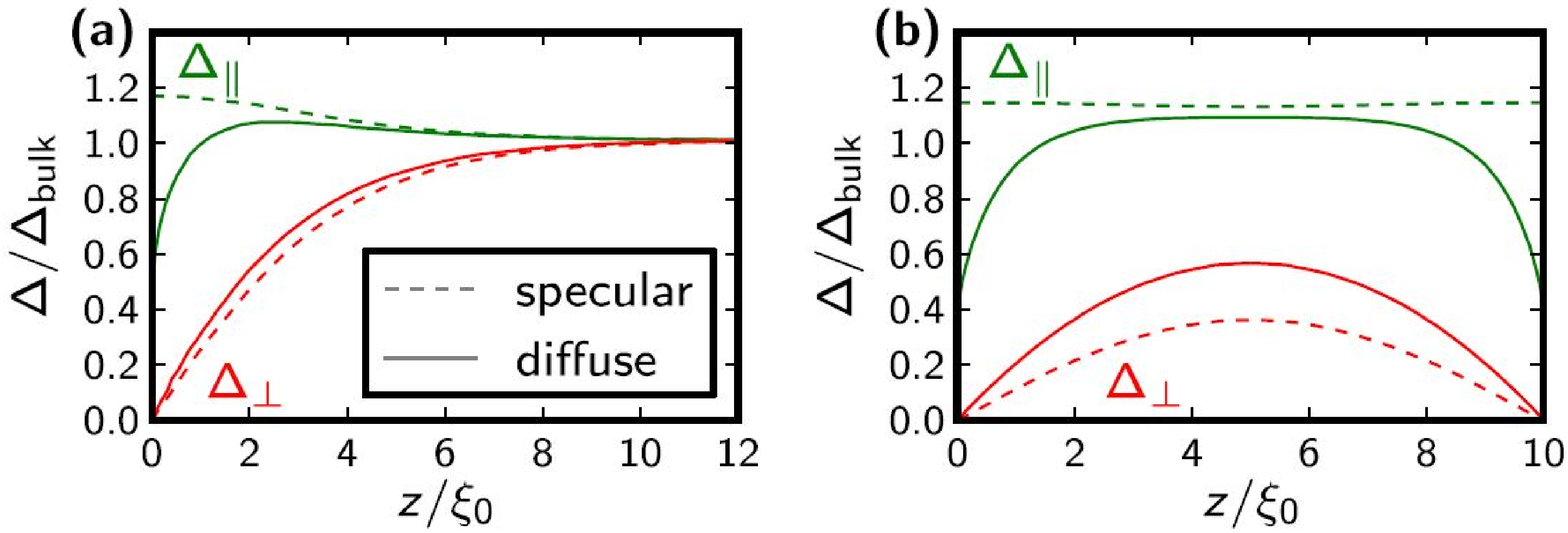}
\end{center}
\caption{Suppression of components of the B-phase gap for specular and diffuse scattering limits \cite{Nagato2000,Nagai2008}. (a) Single surface, (b) Slab of height 10 $\xi_0$. Under such confinement the surfaces, and surface scattering play a dominant role. This provides a laboratory to investigate emergent surface excitations.}
\label{fig:fig-10}
\end{figure} 

\subsection{3.5 Spatially-modulated superfluid order}

There is widespread interest in superfluid/superconducting states in which the order parameter spontaneously acquires a spatial modulation. Such FFLO states~\cite{Fulde1964,Larkin1965} have been widely discussed in spin-singlet superconductors, induced by a spin splitting of the Fermi surface by magnetic fields close to the Pauli limiting field, under conditions where orbital effects are inhibited. This conditions are potentially achieved in heavy fermion superconductors~\cite{Matsuda2007,Kim2016}, or low dimensional organic superconductors \cite{Mayaffre2014}.
\begin{figure}[h]
\begin{center}
\includegraphics[width=3in]{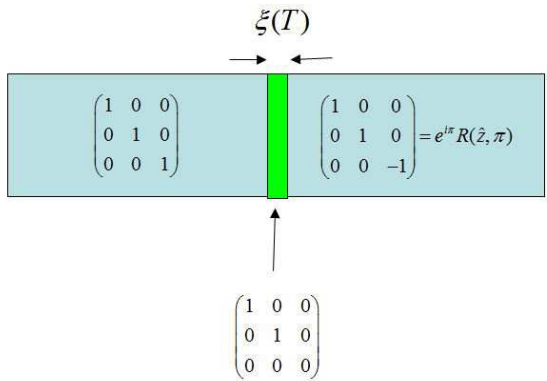}
\end{center}
\caption{The order parameter of the B-phase supports a number of possible ``hard'' domain walls of thickness of order the coherence length \cite{Salomaa1988}. A \emph{simplified} version of one of these is shown. One component of the order parameter changes sign, and the domain wall itself is the planar phase. Under confinement such a domain walls reduce surface pair breaking and can acquire negative surface energy. This is predicted to lead to their proliferation and the formation of the stripe phase, a spatially-modulated superfluid \cite{Vorontsov2007}.}
\label{fig:fig-11}
\end{figure}
A spatially modulated superfluid state has also been predicted in topological superfluid $^3$He-B confined in a thin cavity with slab geometry: the stripe phase \cite{Vorontsov2007}. Since the superfluid is a clean condensed matter system, totally free from impurities, conditions are favourable to observe a state that is \emph{intrinsically} inhomogeneous. A distinguishing feature of this spatially modulated superfluid state is that it arises in a p-wave superfluid with spin-triplet pairing, unlike the spin-singlet cases discussed above. As we have discussed, the superfluid order parameter necessarily varies across the slab due to the boundary conditions on the upper and lower surfaces. However, in addition, an in-plane inhomogeneity is predicted due to the spontaneous appearance of domain walls, on either side of which one component of the order parameter changes sign. These are ``hard'' domain walls, of thickness of order the superfluid coherence length between energetically equivalent states in the B-phase manifold. They were originally discussed in the context of bulk liquid \cite{Salomaa1988}. In this proposal they can be stabilized under confinement, since the presence of the domain wall reduces surface pair breaking, and the domain wall acquires negative surface energy. Their proliferation potentially leads to the formation of a spatially modulated phase.

The simplest possible structure is a periodic array of regularly spaced and linear domain walls, referred to as the stripe phase. Superfluid $^3$He has the advantage that NMR directly probes the spin degrees of freedom of the Cooper pairs. A stripe phase has $q=0$, which has a clear signature in the tipping angle dependence of the NMR response. Its stability is predicted to be favoured by weak coupling, so an experiment was performed at zero pressure where strong coupling effects should be minimised. We therefore chose a $1.1~\mu$m cavity, which has an A-B transition at zero pressure \cite{Levitin2019}. While the experiment rules out the stripe phase, there is NMR evidence for a spatially modulated superfluid of two-dimensional morphology, characterised by two wavevectors, rather than the single wavevector of the stripe phase. We refer to this as the polka dot phase. It is similar to states discussed in the context of FFLO \cite{Matsuda2007}. In our context it may arise because of a lower nucleation barrier of ``dots'' compared to stripes, which are macroscopic in one dimension.
\begin{figure}[h]
\begin{center}
\includegraphics[width=2in]{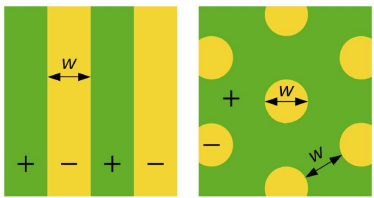}
\end{center}
\caption{Two possible spatially modulations of planar distorted superfluid $^3$He-B under confinement: stripe phase and ``polka-dot''. }
\label{fig:fig-12}
\end{figure}

\subsection{3.6 Nucleation of B-phase under confinement}

In bulk the nucleation of the B phase from the A phase, a first order phase transition, has been widely discussed \cite{Schiffer1995}. Study of the A-B transition under confinement at low pressures, using a torsional pendulum, shows that it is easy to nucleate the B phase, and very little supercooling is observed \cite{Zhelev2017}. We argued that this demonstrates a new intrinsic nucleation mechanism, in which the stripe phase and resonant tunnelling model (significant in models of the early universe) are proposed to play a role~\cite{Zhelev2017}. Using the new techniques of nanoscale confinement, this is subject to further test. We can engineer an isolated ``mesa'' of superfluid $^3$He, cooled through normal ``leads''(where the normal state is stabilized by high confinement). This set-up provides an environment to answer fundamental questions on phase transitions in the superfluid of broad cosmological relevance. 

 \section{4. Topological Mesoscopic Superfluidity: Future prospects}
 \subsection{4.1 Strong confinement}
The demonstration that, when the surface boundary layer includes a thin superfluid $^4$He film, there is no suppression of $T_c$ or the superfluid energy gap, and that surface scattering is therefore specular, opens the way to studies of superfluid $^3$He under progressively stronger confinement. As the ``film'' is made thinner we enter the quasi-2D limit, in which size quantization along $\mathbf z$ plays a role and the Fermi sphere breaks up into Fermi discs, where the number of 2D mini-bands is $j=k_\mathrm F D/\pi$. This opens up a wealth of new quantum states, associated with the integer number of bands, which in principle can be tuned by slab thickness, \cite{Volovik1992b}, including detection the thermal quantum Hall effect in chiral superfluid $^3$He-A (if this is indeed the stable phase).

In our opinion, the quasi-2D regime requires a new theoretical approach on a number of grounds. Firstly the pairing interaction, and strong coupling effects, may be modified, possibly through a dependence of the spin-fluctuation spectrum on dimensionality.  Secondly, quasi-classical theory as presently constituted is not adequate, under conditions of high confinement. This theory was developed to treat the influence of surfaces and interfaces within bulk superfluid $^3$He. The notion of an arbitrary quasiparticle trajectory impinging on the confining surface clearly breaks down in quasi-2D. In this limit the quasi-2D minibands are subject to an effective disorder potential $v(x,y)$ that is determined by the fluctuations in confining cavity height $D + d(x,y)$, due to surface roughness or longer length scale variations in cavity height \cite{Tesanovic1986,Trivedi1988}. Since variations in cavity height can be measured, at least in principle, we have the unusual situation of a disorder potential that can be fully determined experimentally. The success of this approach has already been demonstrated in studies of the flow of an unsaturated normal $^3$He film over a polished silver surface with fully characterised surface roughness \cite{Casey2004b,Sharma2011}. The challenge now is to extend this approach to analyse quasi-2D superfluid $^3$He. [It should be noted that it is necessary to eliminate experimentally other sources of disorder. Each surface should be an equipotential to eliminate fluctuating electric fields, and consequent electro-strictive effects].

The ultimate objective is a 2D p-wave superfluid. Because of the different classes of the topological defects in a p-wave superfluid, the topological phase transitions into this system are particularly rich \cite{Kawamura1999,Korshunov2006}.

\subsection{4.2 Hybrid superfluid structures}
In hybrid metallic nanostructures, normal metals and s-wave superconductors are combined to create new devices. Here a new approach will be the creation and investigation of a range of superfluid $^3$He hybrid nanostructures to reveal edge, surface and interface states in a model topological superfluid. This will serve as a model of such mesoscopic devices based on unconventional, p-wave, superconductors, including topological superconductors. $^3$He allows a degree of control that is unprecedented. The $^3$He order parameter is a $3 \times 3$ matrix with complex components encoding the spin state as a function of position over the Fermi surface, the gap anisotropy and the topology. As we have seen, confinement is the key new control parameter and is used to stabilise distinct superfluid phases or normal phase, and the interfaces between them, with exquisite control over interface quality, the elimination of disorder, and flexibility of geometry.

\subsubsection{4.2.1 Order parameter sculpture} 
The hybrid mesoscopic structure is built from a range of ``materials'', with $^3$He tuned into a particular state by the scale and sculpture of confinement, or the creation of p-wave superfluid meta-materials. For example consider a cavity with a square array of posts separated by $\sim10\xi_0$, for which the phase diagram has been calculated~\cite{Wiman2013}. Near $T_\mathrm c$ the polar phase is stabilized. This geometry should allow the creation of polar phase in the absence of disorder, and even to engineer disorder in a controlled way. [It has recently been shown that the polar phase is stabilized in nematic aerogels, and furthermore stabilizes half-quantum-vortices~\cite{Autti2016}]. At lower temperatures a new variant of the B-phase with four-fold symmetry is predicted~\cite{Wiman2013}.
\begin{figure}[h]
\begin{center}
\includegraphics[width=1in]{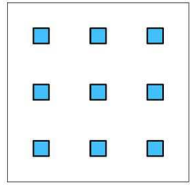}
\end{center}
\caption{Periodic array of posts to stabilize polar phase, and a variant of B-phase with four-fold symmetry \cite{Wiman2013}.}
\label{fig:fig-13}
\end{figure} 

On the other hand in narrow channels the following sequence of phases is predicted~\cite{Wiman2015}: chiral-A; chiral/polar with periodic domain walls; polar.

\subsubsection{4.2.2 Superfluid $^3$He meta-materials}
It is also possible to imagine the creation of superfluid $^3$He meta-materials, such as a regular array of islands, each hosting a macroscopic quantum state, with interconnecting channels. Of interest is a periodic array of islands, which are not equivalent in terms of their superfluid phase (A, B, polar, spatially modulated etc). Furthermore, the interconnecting channels, ``bonds'', can be tuned. A higher level structure in which two or more meta-materials are coupled together is even possible.
 \begin{figure}[h]
\begin{center}
\includegraphics[width=1.5in]{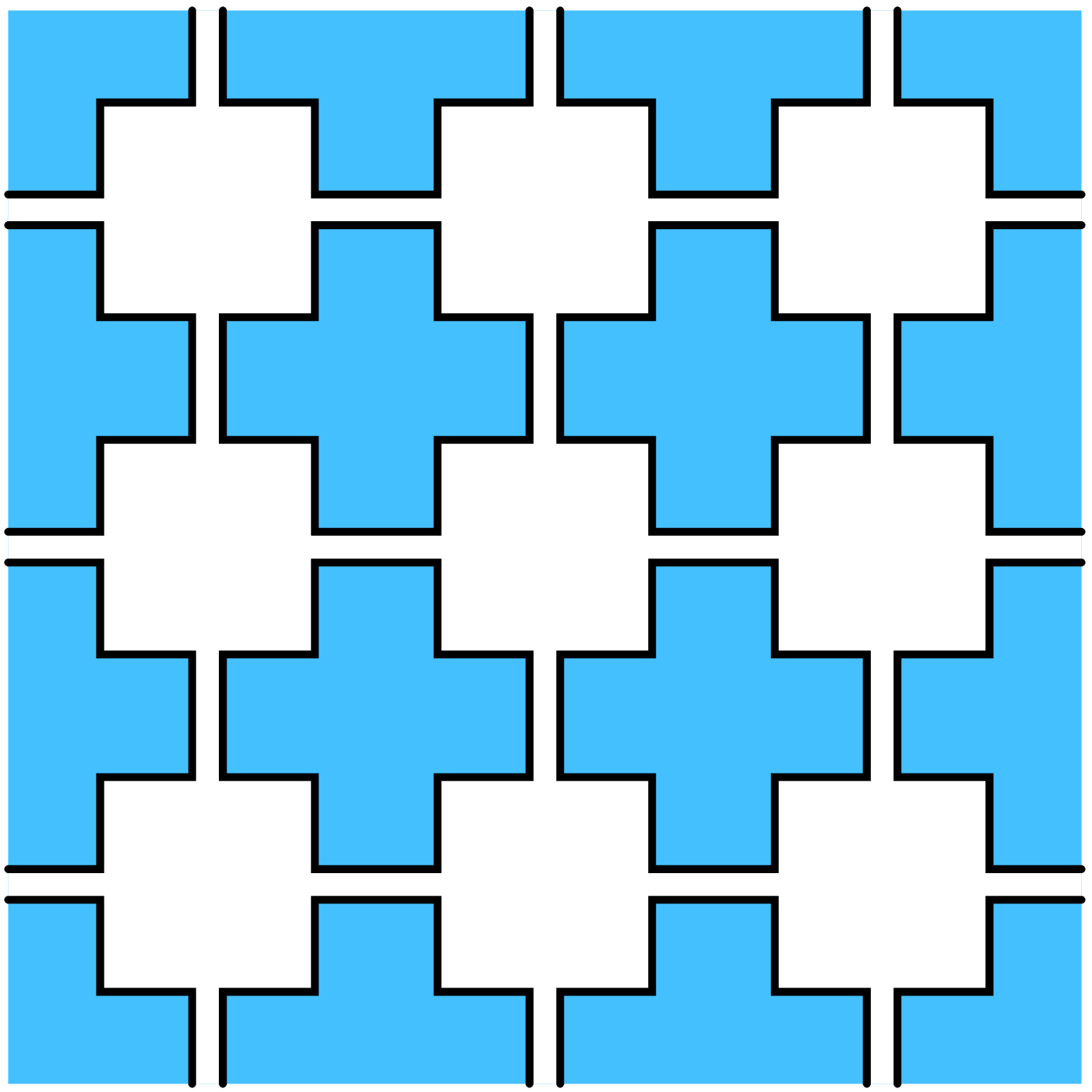}
\end{center}
\caption{Periodic array of inter-connected cavities to create superfluid $^3$He meta-materials. Adjusting height of cavity will stabilize different phases. Not all cavities need be the same height.}
\label{fig:fig-14}
\end{figure} 

\subsection{4.3 Majorana and Majorana-Weyl excitations at surfaces, interfaces and edges}

Topological superfluids/superconductors necessarily host surface and edge excitations. The spectrum of these emergent surface excitations is calculated self consistently with the gap suppression \cite{Nagai2008,Sauls2011,Vorontsov2003}. The energy density of states of these mid-gap excitations depends on the surface scattering conditions. As we have seen a variety of p-wave superfluids are stabilized by confinement. Each superfluid phase has different symmetry, topology and surface/edge excitations.

The time-reversal-invariant B-phase with specular scattering has linearly dispersing Majorana surface excitations~\cite{Murakawa2011}, with strong spin-orbit locking~\cite{Wu2013}. They carry a ground state spin current. Again, these spins are the $^3$He nuclear spins, and there should therefore be clear signatures in NMR response~\cite{Silaev2011}. The quasiclassical theory of how the spin-orbit locked surface excitations may influence the NMR response of the confined superfluid is currently under development~\cite{Silaev2018}. Ultimately we might hope to detect a non-local response of surface Majorana excitations in confined superfluid $^3$He-B using local NMR probes.  

 Furthermore the new topological mesoscopic $^3$He structures offer the prospect of further control over excitations at engineered intra-fluid interfaces. To give an example, we envisage a structure with a step in cavity height at which we can stabilize an interface, either between superfluid and normal liquid (SN) or between two topologically distinct superfluid phases (SS$'$). Unlike an SN interface between a superconductor and normal metal this interface is within a single material (reminiscent of the high quality pn junction in silicon). By combining such ingredients we can create progressively more complex hybrid structures.
\begin{figure}[h]
\begin{center}
\includegraphics[width=4in]{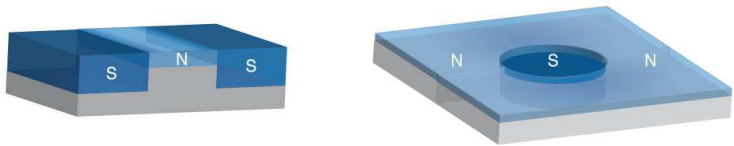} 
\end{center}
\caption{Mesoscopic structures: SNS junction ({SS$'$S} is also possible); superfluid mesa, isolated by normal slabs.}
\label{fig:fig-15}
\end{figure} 
The new feature of such junctions is that we are dealing with a p-wave superfluid. If B-phase, Majorana-like excitations will exist at each SN interface. The spectrum of Andreev bound states in the junction is controlled by the length of the junction $L_\mathrm N$, and we can exploit the fact that at $T_\mathrm c$ the inelastic scattering length is of order 50\,microns.

\subsection{4.4 Further probes of surface excitations}
Furthermore, the breaking of gauge and rotational symmetries in superfluid $^3$He gives rise to gapped Anderson-Higgs modes, as well as gapless Nambu-Goldstone modes. These modes couple to longitudinal and transverse zero sound. Thus, uniquely, superfluid $^3$He offers the prospect to study the coupling of bosonic collective modes of the superfluid order to fermionic degrees of freedom, in particular the predicted exotic surface states~\cite{MizushimaSauls2018}. Moreover, since our topological "superconductor" is a liquid, we can insert probes into it. A new generation of mesoscopic devices, such as nanowires of diameter comparable to the superfluid coherence length is under development, with resonant frequencies comparable to the superfluid gap, to study the interplay of vibration and surface/edge excitations. In part this is motivated by closely related studies of superfluid $^3$He using ions, trapped under the free surface~\cite{Ikegami2013b}.

\section{5. Conclusion}
The cold atoms $^3$He and $^4$He can be grown in atomically layered thin films, on graphite with a variety of pre-platings, or confined in precisely engineered nanofluidic geometries. By manipulating helium in this way it is possible to effectively create new ``materials" with interesting classes of quantum ground states, some anticipated, others emergent. These new materials can be well-characterized, while disorder and impurity effects are small and controllable. In this way important classes of quantum materials can be realized using helium as a model system, exploiting its relative simplicity, and providing a robust confrontation between theory and experiment.
\section{Acknowledgements}
The recent experiments at Royal Holloway reported here were a collaboration with Frank Arnold, Andrew Casey, Brian Cowan, Petri Heikkinen, Lev Levitin, Jan Ny\'eki, Xavier Rojas, Alexander Waterworth, and Jeevak Parpia at Cornell (NSF DMR 17808341). Thanks to Brian Cowan, who helped with the preparation of this manuscript. Discussions with Matthias Eschrig, Piers Coleman, Andrew Ho, Derek Lee, James Sauls are gratefully acknowledged. The research on atomically layered films was most recently supported by EPSRC (UK) through EP/H048375/1. The work on topological superfluidity was most recently supported by EP/J022004/1. We also acknowledge support of the European Microkelvin Platform, and invaluable discussions with its members.





\end{document}